%% file: main.tex
\RequirePackage{fix-cm}
\documentclass[smallextended]{svjour3}       % onecolumn (second format)
\smartqed  % flush right qed marks, e.g. at end of proof
\usepackage{graphicx}
\usepackage{amsmath, amssymb}
\usepackage{bm}
\usepackage{url}
\usepackage{nicefrac}
\usepackage{subfig}
\usepackage{natbib}
\newcommand{\hide}[1]{}

\usepackage[dvipsnames]{xcolor}
\usepackage[normalem]{ulem}
\newcommand{\R}{\mathbb{R}}
\newcommand{\N}{\mathbb{N}}
\newcommand{\NS}{f({\ES})}   % nondominated set
\newcommand{\ES}{X_{E}}   % efficient set
\newcommand{\ve}{\varepsilon}
\newcommand{\itk}{i=1,\dots,m}

\usepackage{algorithm}
\usepackage{algpseudocode}
\usepackage{paralist}
\newcommand{\lb}{l}
\newcommand{\ub}{u}
\newcommand{\lbs}{L}
\newcommand{\ubs}{U}

\DeclareMathOperator*{\argmax}{argmax}

\newcommand{\kd}[1]{\color{black}#1 \color{black}}
\newcommand{\rg}[1]{\color{black}#1 \color{black}}
\newcommand{\el}[1]{\color{black}#1 \color{black}}
\newcommand{\jw}[1]{{\color{black}#1}}
\newcommand{\fs}[1]{{\color{black}#1}}

\journalname{submitted for publication %OR Spectrum
}

\usepackage{todonotes}

\begin{document}

%\input{revision}
%\newpage
%\nonumber
%\setcounter{page}{1}

\title{Multicriteria asset allocation in practice
%\thanks{}
}
% Grants or other notes about the article that should go on the front
% page should be placed within the \thanks{} command in the title
% (and the %-sign in front of \thanks{} should be deleted)
%
% General acknowledgments should be placed at the end of the article.


%\titlerunning{SAA Version 30.04.2020}        % if too long for running head

\author{Kerstin D{\"a}chert \and
        Ria Grindel \and
        Elisabeth Leoff \and
        Jonas Mahnkopp \and 
        Florian Schirra \and
        J{\"o}rg Wenzel
}


\institute{
Kerstin D{\"a}chert$^*$ \and Ria Grindel \and
Elisabeth Leoff \and  %Jonas Mahnkopp \and 
Florian Schirra \and J{\"o}rg Wenzel
\at
Fraunhofer ITWM, Department of Financial Mathematics, Fraunhofer-Platz 1, 67663 Kaiserslautern, Germany\\
%              Tel.: +123-45-678910\\
%              Fax: +123-45-678910\\
\email{firstname.lastname@itwm.fraunhofer.de} \\
$^*$Corresponding author  
%           Jonas Mahnkopp \\
%\email{jonas@mahnkopp.de}
}

\date{\today}

\maketitle

\begin{abstract}
In this paper we consider the strategic asset allocation of an insurance company.
This task can be seen as a special case of portfolio optimization. 
In the 1950s, Markowitz proposed to formulate portfolio optimization as a bicriteria optimization problem considering risk and return as objectives.
However, recent developments in the field of insurance require four and more objectives to be considered, among them the so-called solvency ratio that stems from the Solvency II directive of the European Union issued in 2009. 
Moreover, the distance to the current portfolio plays an important role. 
While literature on portfolio optimization with three objectives is already scarce, applications with four and more objectives have not yet been solved so far by multi-objective approaches based on scalarizations.
However, recent algorithmic improvements in the field of exact multi-objective methods allow the incorporation of many objectives and the generation of well-spread representations within few iterations.
We describe the implementation of such an algorithm for a strategic asset allocation with four objective functions and demonstrate its usefulness for the practitioner. 
Our approach is in operative use in a German insurance company. Our partners report a significant improvement in their decision making process since, due to the proper integration of the new objectives, the software proposes portfolios of much better quality than before within short running time.


% MSC 2010: 90C26 Nonconvex, 90C29 Multicrit, 90C30 Nonlinear
\end{abstract}

\section{Introduction} \label{sec_intro}

Insurance companies have to manage and invest large amounts of money, both from their equity and from premia paid by customers. Investing this capital efficiently is one of the most important challenges insurers face today. Finding a good or optimal investment strategy is a difficult task in itself, and it is even more challenging in a strongly regulated industry such as (life) insurance. 
Investment strategies have to be chosen with various issues in mind, such as the insurer's long-term liabilities, the regulatory environment, different kinds of investment risk and other portfolio properties. 

Today's low-interest rate environment is a challenge for many investors, but especially for life insurers: 
They have to fulfill many old contracts with guaranteed interest rates that are very high compared to the current rates offered at the market.
It is not possible to generate the revenues needed for these liabilities by investing in low-risk assets only. At the same time, the Solvency II directive, introduced by the European Union\footnote{The Solvency II requirements are defined in the directive 2009/138/EC of the European Parliament, in the delegated act from 10 October 2014 and binding technical standards. They are supplemented with supervisory guidelines and recommendations by the national regulators (BaFin in Germany) and the European regulator EIOPA.} in the aftermath of the financial crisis to strengthen the financial stability of the insurance sector, stipulates higher capital requirements for investment in high-risk assets.

Since Solvency~II taking effect, insurers have to calculate their own funds and risks in a standardized manner to prove that their own funds are sufficient to cover their risks in the event of losses: The \emph{Solvency Capital Requirement (SCR)} is calculated to ensure that the insurance company will be solvent over the next 12 months with a probability of at least 99.5\%.
\kd{To achieve this requirement, practitioners typically formulate a minimum value for the solvency ratio of a portfolio. However, they would prefer portfolios with higher solvency ratios to those with smaller ones. Hence, instead of incorporating solvency as a constraint, it should rather be treated as an objective to be maximized.}

In this application-driven paper we consider portfolio optimization with \kd{certain classic} objectives as
well as new objectives like \kd{the solvency ratio.} 
In the following we give a short overview on the vast literature on portfolio optimization. We also discuss portfolio optimization in the light of \el{Solvency~II requirements and} multi-objective optimization.

%--------------------------------
%
%--------------------------------
\paragraph{Portfolio Optimization}
The problem of portfolio optimization has been studied extensively and in many different contexts. The first and foremost goal in a typical portfolio optimization setting is to maximize either the expected utility of the return or the expected return directly.

The classical approach using the concept of utility is often formulated as a constrained maximization problem: The investor chooses a utility function, an increasing function that assigns a subjective value to his or her absolute wealth, which is typically concave due to risk aversion. The goal is then to maximize the expected value of this utility via finding the optimal \el{admissible} trading strategy.%\elout{ in the set of all possible admissible strategies.}

Another fundamental approach is to choose a measure of risk and directly maximize the expected return, now constrained by the amount of risk the investor is willing to accept. This method is related to the modern portfolio theory pioneered by \citet{Markowitz1952}. 

In continuous time, the problem of finding the trading strategy that optimizes expected utility is often called Merton's portfolio problem. Its solution is the famous Merton fraction (\cite{Merton1969}). 
It has since been extended to many more generalized settings such as including trading costs, bankruptcy or non-constant asset parameters (\cite{Karatzas1986}, \cite{DavisNorman1990}, \cite{ShreveSoner1994}, \cite{Korn1998}).

Further approaches include, among a host of other concepts, 
robust portfolio optimization \citep{KimKimFab2014}, 
\el{regime-switching models (\cite{BR04}, \cite{HS04a}, \cite{KLS18}} or worst-case portfolio optimization (\cite{KornWilmott02}, \cite{Seifried10}, \cite{KornLeoff2019}). 
An overview of practical challenges and future trends is given in \cite{Kolm2014}.

The effects of the Solvency~II directive on optimal portfolios have also been considered \el{in the literature recently, using  different settings and concepts.}
\el{\cite{Braun2015} consider Solvency~II requirements in a constrained portfolio optimization framework for an endogenously given amount of equity capital. 
In \cite{Kouwenberg2018}, the author considers a static portfolio optimization problem, where the insurance company wants to maximize the expected return on its own funds. 
\cite{Escobar2019} investigate the implications of the market risk module of Solvency~II on investment strategies in an expected utility framework. 
In all these approaches, the SCR is used as a constraint.
}

%--------------------------------
%
%--------------------------------
\paragraph{Multi-objective Portfolio Optimization}

In this paper we 
%focus on the extension of 
extend Markowitz' bicriteria portfolio optimization problem to more than two and, in particular, more than three objective functions.
As mentioned in \cite{Qi2017} the incorporation of further objectives is not standard yet, however, in the last years there has been growing interest 
%\kdout{at least} 
in incorporating 
%\kdout{one} 
additional criteria
as, e.g., dividends, liquidity or social responsibility.
\cite{Hirschberger2013} present an algorithm that generates the nondominated set of a tricriteria problem that is all linear besides one of the minimized objectives being convex. 
\cite{Koeksalan2016} consider the three objectives \emph{expected return}, \emph{conditional Value at Risk} and \emph{liquidity} in a multi-period stochastic problem. 
Portfolios are generated with the help of an augmented weighted Tchebycheff program.  
\cite{Xidonas2018} focus on a practical decision support tool that is able to deal with multiple objectives. In their empirical testing with data from Eurostoxx 50, they consider the three objectives 
\emph{capital return}, 
\emph{MAD} (mean-absolute deviation% 
) and 
\emph{dividend yield}. 
The Pareto optimal solutions are generated by a set of $\varepsilon$-constraint scalarizations whose right-hand side values are chosen from a two-dimensional grid that is defined in the beginning of the algorithm. 

%--------------------------------
%
%--------------------------------
\paragraph{Our contribution} 

In our \kd{application} we consider a portfolio optimization setting where the aim is to decide on next year's target portfolio. 
The novelty of our approach is the incorporation of four 
criteria into a classic portfolio optimization problem
\kd{and the use of a new efficient method to generate meaningful portfolios for the practitioner.} 
Apart from the classic objectives \emph{return} and \emph{volatility}, we consider the \kd{\emph{solvency ratio}} as well as the \kd{\emph{distance to the current (last year's) portfolio}} as third and fourth objective. %, respectively. 
\kd{While the distance to the current portfolio is meant as a proxy to minimize the transaction volume, 
a well-known goal in portfolio optimization, the maximization of the solvency ratio 
has, to the best of our knowledge, not been treated as an objective function yet.
Since the solvency ratio becomes more and more important for insurers, our approach helps to identify portfolios of high practical relevance. }

The rest of the paper is structured as follows. 
Section~\ref{sec_mco} contains the required basics from multi-objective optimization and explains the applied algorithm in more detail, including the discussion of methodological improvements with respect to a recent approach in multi-objective portfolio optimization.   
In Section~\ref{sec_model} we introduce and discuss the considered model, including all objectives and constraints.
In Section~\ref{sec_impl} the algorithm is applied \kd{to} a real-world use case with four criteria. 
Section~\ref{sec_conclusion} contains the conclusion and further ideas. 

%-------------------------
% Section Intro Multicrit
%
\input{intromulticrit}

%
%
%-------------------------

%--------------------------------
%
%--------------------------------
\section{Model setting, objective functions and constraints}
\label{sec_model}

In this paper, we consider an asset model that differentiates several asset classes such as equity, government and corporate debt, private equity, real estate and a cash position. Some of these asset classes may be further divided according to regional (international, German, emerging markets) or %\jwout{size}
\jw{capitalization} (large cap, medium cap, small cap) aspects. Under one such asset class (e.g., German large cap equity) we usually subsume several investments (such as shares in Daimler, BASF, etc.) and consider them identical. In our case, this leads to 13 asset classes, but we will more generally assume $n$ asset classes.

\jw{Asset class $i$ is characterized by its \emph{expected annual return} $\mu_i$, and the \emph{expected variance} $\sigma_i^2$ of its return, that is the expected squared deviation of $\mu_i$ from the true annual return. Moreover,} \jw{different asset classes $i$ and $j$ are related by the \emph{expected covariance}} \rg{$\sigma_i \sigma_j \rho_{ij}$,} \jw{that is the expected product of the deviation of $\mu_i$ and $\mu_j$ from the true annual returns. 
\kd{The parameter $\rho_{ij}$ is called \emph{correlation}.}
All these characteristics can, e.g., be estimated from historical time series and be adjusted by expert knowledge. For our purpose, we consider these numbers as given.}

\jw{We want to construct a portfolio of these asset classes that} satisfies certain conditions at the investment horizon $T$.
\jw{We denote by $\omega_i$ the (current) \emph{weight} of asset class $i$ in this portfolio, e.g., the proportion of today's value of asset class $i$ to the value of the portfolio, and by $\omega=(\omega_1,\dots,\omega_n)$ the \emph{weight vector}.} 
In particular, the expected \jw{annual} return 
of the portfolio
is given as
\begin{equation}\label{eq:mu}
    \mu\jw{(\omega)} 
    = \sum_{i=1}^n \omega_i \mu_i.
\end{equation}
Similarly, \jw{we consider}
the \kd{volatility $\sigma$}
of the portfolio\jw{, which} is given as
\begin{equation}\label{eq:sigma}
    \sigma\jw{(\omega)} := \sqrt{\sum_{i,j=1}^n \omega_i \omega_j \sigma_i \sigma_j \rho_{ij}}.
\end{equation} 

A typical portfolio of asset classes is given in Table~\ref{tab:assetclasses}. We will use this portfolio as a starting point throughout our examples.

\begin{table}
\caption{Example asset classes with their current weights $\omega_\jw{i}$ \jw{in the portfolio}, returns $\mu_\jw{i}$ and volatilities $\sigma_\jw{i}$ \jw{for $i=1,\dots,n$}.}
    \label{tab:assetclasses}
    \begin{tabular}{l|r|r|r}
    \hline\noalign{\smallskip}
        \multicolumn{1}{c|}{Asset class $i$} & 
        \multicolumn{1}{c|}{$\omega_i$} & 
        \multicolumn{1}{c|}{$\mu_i$} & 
        \multicolumn{1}{c}{$\sigma_i$} \\
\noalign{\smallskip}\hline\noalign{\smallskip}
        Real Estate Germany & 5.87\% & 5.30\% & 13.00\%\\
        Real Estate Intl.\ & 4.99\% & 6.00\% & 14.00\%\\
        Equity Intl.\ Large Cap & 6.22\% & 6.50\% & 11.18\%\\
        Equity Germany\ Large Cap & 12.74\% & 5.57\% & 14.10\%\\
        Equity Intl.\ Small Cap & 4.32\% & 5.95\% & 12.72\%\\
        Emerging Markets Equities & 8.52\% & 8.00\% & 13.00\%\\
        Private Equity  & 3.51\% & 8.50\% & 18.00\%\\
        Government Debt & 19.45\% & 0.30\% & 4.00\%\\
        Corporate Debt & 14.83\% & 1.00\% & 3.60\%\\
        Infrastructure Finance & 0.50\% & 3.20\% & 5.70\%\\
        Fixed Income & 5.33\% & 0.40\% & 2.50\%\\
        Asset Backed Securities & 7.74\% & 0.30\% & 2.10\%\\
        Cash & 5.98\% & 0.00\% & 0.00\% \\ 
        \noalign{\smallskip}\hline
    \end{tabular}
\end{table}

We assume \jw{a one-period model, i.e.,}
we can instantly rebalance our investments so that a proportion of $\omega_i$ is invested in asset class $i$ and
the expected returns $\mu_i$, 
\kd{variances $\sigma_i^2$,}
and correlations $\rho_{ij}$ of the asset classes remain constant throughout the investment horizon.

\subsection{Objective functions}

\paragraph{Return and volatility}
Following the concept of Markowitz, \kd{we consider return and volatility} 
as given in \eqref{eq:mu} and \eqref{eq:sigma}.
There are no assumptions on the investor’s preferences such as a specific form of utility function. We only make the natural and standard assumption that the insurer prefers higher expected return and lower volatility. Thus, return is maximized while volatility is minimized.

%-------------------------------------------
%
\input{solvency}
%
%------------------------------------------

\paragraph{Distance to the current portfolio}
%\paragraph{Distance to a reference portfolio}
In the application at hand we consider a one-period model, \kd{determined once a year.
Since the input data changes from year to year, there is a need to determine a new portfolio every year. However, due to transaction costs, the insurers favor new portfolios which do not deviate too much from last year's portfolio. 
In the literature, there are different ways to model transaction costs, in particular very sophisticated ones involving discrete variables, which, however, turn the problem into a mixed-integer optimization problem. In order to keep the problem continuous, we use the distance to the current portfolio measured by an $l_1$-norm here. 
} 
\kd{
Let $\overline{\omega} \in [0,1]^n$ be the weights of the current portfolio. To find a portfolio with minimal distance to this portfolio we minimize the $l_1$-norm 
\begin{equation} \label{eq:distance}
 \lVert \omega - \overline{\omega} \rVert_1 =
 \sum_{i=1}^{n} | \, \omega_i - \overline{\omega}_i |.
\end{equation} 
Due to the absolute values, the objective function is not differentiable everywhere. It is well known that differentiability can be achieved by a reformulation of the absolute values with the help of artificial variables and additional inequalities. However, nowadays most solvers can handle these sorts of functions directly. 

Besides using last year's portfolio we can also think of other special portfolios a user wants to relate to. Therefore, in the following, we use the notion reference portfolio, in order to emphasize that any portfolio could be chosen instead. 
}

\subsection{Constraints}
We impose the standard assumption that all portfolio weights sum up to $1$. 
We also assume that all weights are non-negative, i.e., we do not allow for short selling.

\kd{Optionally, the user can restrict the proportion of assets further by indicating lower and upper bounds. 
Besides, it is also possible to specify lower and upper bounds for so-called asset groups which are a set of certain assets, e.g., shares or real estate.}

\kd{
\subsection{Problem formulation}
We can now concisely state our four-criteria optimization problem with objective functions \eqref{eq:mu}, \eqref{eq:sigma}, \eqref{eq:solvency} and \eqref{eq:distance} and the constraints described above. 
The overall problem reads
\begin{align}
    & \max \, \mu(\omega) \notag \\[.1cm] \notag
& \min \, \sigma(\omega) \\[.1cm] \notag
& \max f_\mathrm{solvencyratio}(\omega) \\[.1cm] \label{eq:prob} \tag{MOP}
& \min \, \lVert  \omega - \overline{\omega} \rVert_1 %\sum_{i=1}^{n} | \omega_i - \overline{\omega}_i | 
\\ \notag   
& s.t.\ \\\notag
& \quad  \sum_{i=1}^n \omega_i = 1 \\\notag
& \quad l_{I_g} \leq \sum_{i \in I_g} \omega_i \leq u_{I_g} \quad \forall \, g=1,\dots,G \\\notag
& \quad \omega_i \geq 0 \quad \forall \, i=1,\dots,n 
\end{align}
where $I_g \subset \{1,\dots,n\}$, $g=1,\dots,G$, $G \in \N$, 
is a so-called asset group consisting of a subset of the given assets. The sum of the weights in asset group $I_g$ is bounded by   
$l_{I_g}, u_{I_g} \in (0,1)$. 
If $|I_g|=1$, i.e., if the asset group $I_g$ contains only one asset, the constraint models lower and upper bounds of one specific asset. 
}

%-----------------------------------
%
\input{application}
%
%------------------------------------

\subsection{Advantages from a practical point of view}

The process of determining next year's portfolio can be a challenging task for a big company with many stakeholders to consider. 
Expectations of different parties with possibly contradicting interests have to be integrated and met as best as possible. 
Additionally, 
regulations for %, e.g., 
banks and insurance companies have grown considerably over the last ten to twenty years, which induce further limitations on the portfolio choice. 
This leads to a need for more complex procedures to choose a portfolio which satisfies all regulatory requirements as well as the expectations of the stake- and shareholders. 

However, current procedures %like these 
can not respond \jw{appropriately} to these new requests.
Typically, investors either use a bicriteria Markowitz approach and try to meet the other criteria by some heuristic approach or they apply brute force methods. 
Interesting possible variants are collected, evaluated and then filtered to get a final portfolio. Both ideas are not satisfying and 
tend to be time-consuming and demanding for the 
underlying decision processes.
Multicriteria approaches are able to better cope with the new requests compared to traditional techniques. In particular, the following aspects have been reported as \jw{being} useful by the investors.

\paragraph{Gain in objectivity and time savings}
Because all \jw{relevant} criteria can be integrated in the model, a common basis exists \jw{satisfying} the standards of the different portfolio managers\jw{. Hence,}\ less heuristic approaches need to be used to identify the most preferred portfolio. 
The impact of a change in portfolio \jw{allocation} on the various criteria is \jw{immediately} visible and can be \jw{taken into account directly.} 
This leads to a \jw{much more transparent} portfolio \jw{allocation} process \jw{and} results in considerable time-savings during the whole decision process. 
The final portfolio selection is \jw{substantiated} by actual optimization rather than intuition. 
This also helps different investors to \jw{better agree on a portfolio allocation} that is commonly accepted.

\paragraph{Better detailing by \jw{user} interaction}
The sliders together with the possibility to narrow the search space allow the investor to inspect parts of the outcome space that are of particular interest more closely. 
While this concept is rather standard in interactive multi-objective optimization, it is not present in currently used decision support tools of insurance companies. Hence, compared to other procedures, this feature \jw{offers} the new 
ability to better fine-tune trade-offs. 

\section{Conclusion}
\label{sec_conclusion}

In this paper we apply a recent multi-objective optimization algorithm based on Tchebycheff scalarizations to a real-world portfolio optimization problem with four objectives. 
More precisely, we tackle the problem known as strategic asset allocation in the context of insurance companies. 
Apart from the classic objectives of maximizing return and minimizing risk, a solvency ratio is maximized and the distance to a specified portfolio is minimized. 
The incorporation of these additional objectives allows the generation of portfolios that are much closer to the expectations of the involved investors compared to portfolios generated with other approaches.
The described concept has led to a decision support tool that is in operative use in a German insurance company. 
The tool is flexible and allows the incorporation of further objectives. New challenges include questions on how to present results with many objectives to the investors and how to support them even more to find a final compromise decision.

\begin{acknowledgements}
We thank our industrial partner for the fruitful collaboration and especially for the trust to install a completely new solution concept based on multi-objective optimization. 
\end{acknowledgements}

\bibliographystyle{spbasic}      % basic style, author-year citations
\bibliography{references}   % name your BibTeX data base

\end{document}

%% file: intromulticrit.tex
%!TEX root = main.tex
\kd{\section{Multi-objective and Markowitz portfolio optimization} \label{sec_mco}}
%\section{Multi-objective optimization} \label{sec_mco}

In this section we first introduce common notions in multi-objective optimization. Then we speak about how to solve these problems. 

\kd{Let us consider the general form 
\begin{equation}\label{chap1:optprob}
\min_{x \in X} f(x)=(f_1(x),\dots,f_m(x))^{\top} 
\end{equation}
of a multi-objective optimization problem 
with \emph{feasible set} $X \subseteq \R^n$ 
and 
with $m \geq 2$ \emph{objective functions} $f_i:X \to \R, \, i=1,\dots,m$. 
We assume that the functions $f_{i}, i=1,\dots, m$, are continuous and that 
$X$ is non-empty and compact. 
The image of the feasible set is denoted by $f(X)\subseteq \R^m$.  
}

When dealing with optimization problems with more than one objective we cannot expect to compute ``an optimal solution'' characterized by a solution that has the globally or locally smallest objective function value. 
Instead the single-objective concept of \emph{optimality} is replaced by the concept of \emph{Pareto optimality} (also called \emph{efficiency}). 
\kd{
\begin{definition}[Pareto Optimality / Efficiency]
\label{def_Pareto_eff}
A feasible solution $x \in X$ is called \emph{Pareto optimal} or \emph{efficient} if there is no $\hat{x} \in X$ with 
$f_{i}(\hat{x}) \leq f_{i}(x)$ for all $i =1,\dots,m,$ and $f_{j}(\hat{x}) < f_{j}(x)$ for some $j \in \left\{1,..,m \right\}$. 
\end{definition}  
We denote the set of efficient solutions by $\ES$. 
The image set $\NS$ is called \textit{Pareto front} or \textit{nondominated set}, its elements are called \textit{nondominated}.
A slightly weaker concept is the so-called weak Pareto optimality or efficiency.
\begin{definition}[Weak Pareto Optimality / Efficiency]
\label{def_Pareto_weff}
A feasible solution $x \in X$ is called \emph{weakly Pareto optimal} or \emph{weakly efficient} if there is no $\hat{x} \in X$ with $f_{i}(\hat{x}) < f_{i}(x)$ for all  
$i =1,\dots,m$. 
\end{definition}  
}
From a practical perspective, nondominated points are compromises between the conflicting objectives. It is then up to the decision maker to choose a compromise that suits his or her needs best.
%, e.g., a portfolio with a certain balance of risk and return. 

\subsection{Scalarizations} \label{subsec:scal}
How to compute these nondominated points? A common approach to solve multi-objective optimization problems consists in a so-called \emph{scalarization}. 
This means that the vector-valued optimization problem is reformulated to a scalar-valued one which then can be solved with the help of classic single-objective optimization methods. 
The easiest and most common scalarization is the weighted sum approach 
\kd{
\begin{equation} \label{chap1:WS}
\min_{x \in X} \;  \displaystyle \sum_{i=1}^m \lambda_i f_i(x), 
\end{equation}
}
in which each of the multiple objective functions is multiplied by 
\kd{a so-called weight $\lambda_i \geq 0, i=1,\dots,m,$ where $\sum_{i=1}^m \lambda_i=1$.}
By varying the values of the parameters, different solutions can be found. It can be shown that for every positive weight vector, a nondominated point is found. However, only for convex optimization problems it holds true that every nondominated point can be generated for some weight vector. If the problem is either non-convex or convexity can not be guaranteed, other scalarization techniques as, e.g., the $\varepsilon$-constraint method or the weighted Tchebycheff method should be applied \kd{which are described in the following.} 

\kd{
%-------------------------------------- EC
The \emph{$\ve$-constraint method} was first proposed in \cite{haimes71} and is discussed in more detail in \cite{chankong83}. 
In this method, one of the objectives $f_i$ with $i\in\{1,\dots,m\}$ is selected and minimized whereas bounds are imposed on all other objectives, which yields 
\begin{equation}\label{chap1:EC}
\begin{array}{ll} 
\min & f_{i}(x)\\[.1cm]
\text{s.t.} & f_{k}(x) \leq \ve_{k} \quad \forall \, k=1,\dots,m, \, k \neq i,\\[.07cm]
& x \in X,
\end{array}
\end{equation}
where $\ve \in \R^m$. 
%Note that component $i$ of vector $\ve$ is not used in~\eqref{chap1:EC}.
It is well-known that every feasible solution of~\eqref{chap1:EC} is weakly efficient.  If the solution is unique, then it is efficient. 
On the other hand, for every efficient solution $\bar{x} \in \ES$ there exists %an index $i \in \{1,\dots,m\}$ and 
a vector $\ve \in \R^{m}$ such that $\bar{x}$ solves \eqref{chap1:EC} for any $\itk$. 
More precisely, every efficient solution $\bar{x} \in \ES$ is an optimal solution of \eqref{chap1:EC} for any $i=1,\dots,m$ and $\ve=f(\bar{x})$. 

%------------------------------------------ WT
A scalarization with similar theoretical properties is the
\emph{Weighted Tchebycheff method}.
It was introduced in \citet{bowman76} and studied in detail in \citet{steuer83}. 
It is defined as
\begin{equation} \label{eq:WTN}
\min_{x \in X}  \max_{i=1,\dots,m} 
%\left\{ 
w_{i} \cdot \left| f_{i}(x)-z_{i}^{\star} \right| 
%\right\} 
%\jw{ w_{i} \cdot \left| \, f_{i}(x)-z_{i}^{\star} \right|}
\end{equation}
with $w \in \R^m_{>}$ %$w > 0$. 
and $z^{\star} \in \R^m$ a reference point.
If the reference point is chosen 
%as the (global) ideal point or below, 
so that no feasible point lies in the `lower left part' of the reference point, i.e.\ that 
the negative orthant attached to the reference point is empty,
the absolute values can be dropped. 
Moreover, the objective function can be reformulated as  
\begin{equation} \label{eq:RWTN}
\begin{array}{ll} 
\min & t \\
\mbox{s.t.} & t \geq w_i \left( f_i(x) - z_i^{\star} \right), \quad \itk,\\[0.07cm]
& t \in \R, x \in X,
\end{array}
\end{equation}
see \cite{steuer83}.
This formulation is particularly useful when all underlying functions are differentiable since the overall problem becomes differentiable. 
It is well-known %\citep{bowman76} 
that every solution of~\eqref{eq:WTN} and~\eqref{eq:RWTN} is weakly efficient, and efficient if the solution is unique.
Conversely, for every efficient solution $\bar{x} \in \ES$ there is some $z^{\star} \in \R^m$ and $w \in \R^m_{>}$ such that $\bar{x}$ solves~\eqref{eq:WTN}. 

There are ways to assure nondominance instead of only weak nondominance for the $\ve$-constraint and Weighted Tchebycheff method, which are particularly important in the discrete context where the occurrence of weakly nondominated points is rather frequent. 
In the continuous case weak nondominance only appears in non-convex regions where the nondominated set is unconnected, and there only at the `boundary points' of two unconnected parts. Since the portion of these points is rather small in general, 
we %suppose that the outcomes of the scalarizations are nondominated even if we can only prove them to be weakly nondominated. 
do not apply specific methods to enforce nondominance. }

%-----------------------------
%
%-----------------------------
\subsection{Representation} \label{subsec:rep}

Continuous multi-objective optimization problems as the problem at hand have an infinite number of nondominated points. In general, the nondominated set can not be described analytically, thus, a finite set of points in this set is generated instead. This finite set is called \emph{representation} or \emph{approximation} of the nondominated set, where a representation typically consists of nondominated points while an approximation not necessarily does. 

\kd{The approach used in \cite{Xidonas2018} generates a representation by solving a sequence of $\varepsilon$-constraint scalarizations with different right-hand side values. 
Therefore, an equidistant two-dimensional (in general $(m-1)$-dimensional) grid is computed in the beginning of the algorithm based on the ranges of the objective functions. 
While this approach is rather easy to implement, the rigid grid makes this approach 
inflexible in the sense that it cannot adapt to the shape of the Pareto front. 
Typically, some of the scalarized optimization problems are infeasible, some others yield nondominated points that are already known. 
While \cite{Xidonas2018} present certain enhancements like an "early-exit-strategy", they can not completely avoid these undesired effects. 

In contrast, our approach is flexible in the sense that
the solution process constantly adapts to the Pareto front. 
Infeasible problems or multiply generated solutions 
do not appear as long as the invoked single-objective optimization solvers work reliably.
Our approach 
refines in every iteration where it is needed most, i.e.\ where a certain approximation error is maximal. Details are given in the following. 
}

%-----------------------------
%
%-----------------------------
\subsection{Box algorithm} \label{sec:boxalgo}

In our implementation we follow the algorithmic concept of \cite{DaeTei20} which uses a decomposition of the search region into \kd{a set of} (hyper-)boxes \kd{$\mathcal{B}$. 
Each box $B=[l,u]$ is a rectangular set defined by a lower bound $l \in \R^m$ and an upper bound $u \in \R^m$,} where $m$ denotes the number of considered objectives.
At initialization an $m$-dimensional box \kd{$B_0$} is created
\kd{and the set of boxes is initialized by $\mathcal{B} = \{ B_0 \}$.} 
The ranges \kd{$l_0 \in \R^m$ and $u_0 \in \R^m$} of this initial box are obtained by first minimizing every objective individually and then taking the minimum and maximum value with respect to every objective. 
This approach is also known as \emph{Payoff-table} and used, e.g., in \cite{Xidonas2018} as well. 
\kd{In the case of more than two objectives, }
the resulting box does not necessarily contain the entire 
nondominated set but is in most cases sufficient for the decision maker
\kd{who wants to find a portfolio that represents a good balance among all considered objectives.} 
In each of the following iterations, one box is selected for refinement, i.e., a new point in this box is computed. 
The idea is to always pick a box so that a new point is added in a region that is not well represented yet. Therefore, we compute the smallest edge of each box and select the one with the largest value, i.e., 
\kd{we determine 
\begin{equation} \label{eq:boxselect}
\argmax_{B=[l,u] \in \mathcal{B}} \; \min_{i=1,\dots,m} \{ u_{i} - l_{i} \}
\end{equation}
and use the resulting box for further refinement.

As scalarization we use the Weighted Tchebycheff method. The reason to use this scalarization is twofold. First, we can reach non-convex parts of the nondominated set, second we can search the selected box `uniformly', i.e.\ the computed solution most probably lies on the diagonal of the box. This is different to the $\ve$-constraint method 
where priority is given to one objective function and, hence, solutions rather lie at the boundary of the selected box. 
}

The lower and upper bound of the selected box are used to define the parameters of the Weighted Tchebycheff scalarization. 
\kd{
More precisely, we use the lower bound $l$ as the reference point and compute the weights according to \citep{steuer83} by
\begin{equation} \label{eq:tchebweights}
w_{i} = 
\frac{1}{(u_{i} - l_{i}) \cdot \sum_{j=1}^m \frac{1}{(u_{j}-l_{j})}}.
\end{equation}
}
The idea is to move from the lower bound of the box along its diagonal
until a point $f(\bar{x})$ is found. 
\kd{If the point lies in the interior of the selected box, it must be a new (weakly) nondominated point. Otherwise, we can discard the box since it does not contain any new points. The latter case is important in the discrete context but rarely happens in the continuous case. Nevertheless, due to numerical issues, it might happen that the solver does not generate a point in the considered box. Then, the box is removed and another box is chosen. 

Algorithm~\ref{alg:box} shows the general procedure. 
A non-trivial step is hidden in lines~\ref{alg:box:updatelub} and~\ref{alg:box:updatellb} within $\texttt{newUpperBounds}(\ubs, f(\bar{x}))$
and $\texttt{newLowerBounds}(\lbs, s)$, 
which contains the update of the bounds $\lb$ and $\ub$ by which the boxes are defined. 
Procedure~$\texttt{newUpperBounds}(\ubs, f(\bar{x}))$ consists of the following steps: 
First, all $\ub \in \ubs$ have to be detected, for which $f(\bar{x}) < \ub$ holds. From each of these bounds, at most $m$ new bounds $\ub^i, i=1,\dots,m,$ of the form 
\begin{equation}
\ub^i = \left\{
\begin{array}{ll}
f_{k}(\bar{x}), & k=i\\
\ub_{k}, & k \neq i
\end{array}
\right.
\end{equation}
are created and the former bounds $\ub$ are deleted. Due to redundancies not all $m$ child bounds are needed. The necessary bounds can be detected by a criterion before their creation. For details we refer to \cite{DaeKla15}, \cite{KlaLacVan15} and \cite{DaeKlaLacVan17}. 
The update of the lower bounds $\lb \in \lbs$, that contain the current solution, works in a similar fashion. The only difference is that it is beneficial to use the Tchebycheff vertex $s_{i} = l_{i} + t/w_{i}$ instead of $f(x)$ to obtain tighter bounds. For details we refer to \cite{DaeTei20}. 
}

\kd{
\begin{algorithm}[htbp]           
\caption{Box algorithm}  \label{alg:box} 
\begin{algorithmic}[1]
\algblockdefx[Input]{Input}{EndInput}
    {\textbf{input:}}
    {}
\algblockdefx[Output]{Output}{EndOutput}
    {\textbf{output:}}
    {}
\Input
\begin{compactitem}
	\item a multi-objective problem with $m$ objectives
	\item a maximum number of iterations $maxit$
\end{compactitem}
\EndInput
\Output
\State	a set of (weakly) nondominated representation points $Z$
\EndOutput

\algblock[Name]{Start}{End}
\Start
\State $Z \gets \emptyset$
\State Compute initial lower bound $\lb_0$ and upper bound $\ub_0$ from payoff table 
\State $\lbs \gets \left\{\lb_0 \right\}$, $\ubs \gets \left\{\ub_0 \right\}$, $B_{0} = \left[\lb_0,\ub_0 \right]\subseteq \R^{m}$
\State $\mathcal{B} \gets B_{0}$
\State $it = 0$;
\While {$it < maxit$}
		\State Select $B = [\lb,\ub]$ from $\mathcal{B}$ according to \eqref{eq:boxselect}
		\State Compute Tchebycheff weights $w$ according to \eqref{eq:tchebweights} 
		\State Solve \eqref{eq:RWTN} 
		and obtain solution $(t, f(\bar{x}))$
		\State Compute Tchebycheff vertex $s \in \R^{m}$ by $s_{i} = l_{i} + \frac{t}{w_{i}}, 			i=1,\dots,m$ 
		\If {$f(x) < u$}
			\State $Z \gets Z\cup \left\{ f(\bar{x}) \right\} $
			\State $\ubs \gets \texttt{newUpperBounds}(\ubs, f(\bar{x}))$ \label{alg:box:updatelub}
			\State $\lbs \gets \texttt{newLowerBounds}(\lbs, s)$ \label{alg:box:updatellb}
			\State $\mathcal{B} := \left\{B = \left[\lb,\ub\right] | \, \lb \in \lbs, \ub \in \ubs, \lb < \ub \right\}$
			\Else 
			\State Remove $B$ from $\mathcal{B}$
			\EndIf
		\State $it = it + 1$;
\EndWhile
\State \Return $Z$
\End
\end{algorithmic}
\end{algorithm}

\subsection{Markowitz portfolio or mean-variance optimization}
In Section~\ref{subsec:scal} we presented three classical scalarization approaches. 
Indeed, formulations of the mean-variance optimization (MVO) 
can be interpreted as a Weighted Sum or $\ve$-constraint problem.
\cite{Kolm2014} present the MVO as
\begin{equation} \label{mvo:ws}
\max_{\omega \in \Omega} \left( \mu^{\top} \omega - \lambda \omega^{\top} \Sigma \omega \right),
\end{equation}
where $\lambda$ denotes a risk aversion parameter measuring the relative importance between the expected portfolio return $\mu^{\top} \omega$ and the portfolio risk $\omega^{\top} \Sigma \omega$. Details on the notation are given in the next section, here we only want to draw the connection to the scalarizations presented in Section~\ref{subsec:scal}. Formulation~\eqref{mvo:ws} is a Weighted Sum of the two objectives `maximize return' and `minimize risk'.
Alternative formulations of the MVO presented in \cite{Kolm2014} are
\begin{equation} \label{mvo:eq1}
\begin{array}{ll} 
\displaystyle \max_{\omega \in \Omega} & \mu^{\top} \omega  \\
\mbox{s.t.} & \omega^{\top} \Sigma \omega \leq \sigma^2_{max}  
\end{array}
\end{equation}
and
\begin{equation} \label{mvo:eq2}
\begin{array}{ll} 
\displaystyle \min_{\omega \in \Omega} & \omega^{\top} \Sigma \omega   \\
\mbox{s.t.} & \mu^{\top} \omega \geq R_{min},  
\end{array}
\end{equation}
thus, $\ve$-constraint problems. 
The drawback of both $\ve$-constraint formulations is that the solution obtained is typically close to the selected parameter, i.e.\
the achieved risk in \eqref{mvo:eq1} is close to $\sigma^2_{max}$, the obtained return in \eqref{mvo:eq2} is close to $R_{min}$.
By using the Weighted Tchebycheff scalarization, points that are balanced between the considered objectives are achieved, in general, which is the reason why we choose this scalarization within our algorithm. 
}

%% file: solvency.tex
\paragraph{Solvency ratio}

In Section~\ref{sec_intro} we introduced the Solvency II regulations which include among others that insurers have to prove that they have enough funds to secure their risky assets. 
\fs{This is ensured by controlling that the risk to which a portfolio is exposed is always matched by a sufficiently high value of own funds, meaning that the ratio of own funds to the risk must be larger than a given threshold. The crucial part is the calculation of the risk, hence we give a short explanation on how it is done. \jw{A good motivation and derivation (in German) of the SCR formula can be found in \citet[Section~3.5.2.1b, p.~318]{GVK557107601}.}}

Firstly, the allocation decision is used to determine net risks for the market risk defined in 
Solvency~II. 
\fs{These include the}
\kd{following eight risk types: }
\kd{\emph{interest rate up}, \emph{interest rate down}, \emph{equity type 1}, \emph{equity type 2}, \emph{property}, \emph{spread}, \emph{currency up} and \emph{currency down}.}
The net risk represents \jw{the loss} 
in the eight scenarios compared to the most probable scenario (best estimate). Stress parameters assumed in the respective scenario (e.g., the amount of equity losses) are calibrated to the market 
such that the stress corresponds to a 200-year event. 
This means \kd{that}
\jw{the net risk is the difference between own funds and value at risk for a time horizon of one year and a probability $\alpha = 1/200$.}
\jw{By linearization and approximation one can assume that}
for
\fs{a weight} \kd{vector $\omega \in [0,1]^{n}$} 
\jw{the net risk is given as $A\omega +b$, where $A \in \R^{n \times 8}$ and $b \in \R^8$.
This roughly corresponds to the stress definition required by the regulatory authorities, as each asset generates a risk factor. 
We denote the resulting function by
\begin{align*}
  f_{\mathrm{netrisk}} : [0,1]^n &\to \mathbb R^8 \\
  \omega &\mapsto A\omega + b,
\end{align*}}
where \jw{the dimensions $i=1,\dots,8$ correspond to the risk types mentioned above in} \kd{the given}
\jw{order.}
\kd{This order plays a role in the subsequent formulas since the risk types are aggregated differently. First, we build}
\fs{
\begin{align*}
    f_\mathrm{aggregation}:\mathbb{R}^8&\to\mathbb{R}^5 \\
    \jw{\boldsymbol{x}} &\mapsto 
    \left(
    \begin{array}{l}
        \max \{ x_\jw1, x_\jw2 \} \\
        \jw{\sqrt{x_3^2 + 1.5 x_3x_4 + x_4^2}} \\
        x_\jw5 \\
        x_\jw6 \\
        \max \{ x_\jw7, x_\jw8 \}
    \end{array}
    \right)
\end{align*}
}
\kd{(Note that the square root is well defined even in case that either $x_3$ or $x_4$ is negative, since then $x_3^2 + 1.5 x_3x_4 + x_4^2 = (x_3 + x_4)^2-0.5x_3x_4$ is positive.)}

%-------------------
% Market risk
%-------------------
\fs{In the next step, the risk types are aggregated into one risk type, the market risk,} 
\kd{by}
\fs{using \jw{two} correlation matrices and taking the maximum of the two correlation scenarios. 
Note that an additional type of risk is added, the concentration risk, which is considered to be constant in the context of portfolio optimization
\jw{and denoted by $c_1$}. The aggregation function} 
\kd{then reads}
%looks like this:
\fs{
\begin{align*}
    f_\mathrm{market}:\R^{5}&\to\R\kd{_0^+}  \\
    x & \mapsto \sqrt{\max \{ x^{\top} P_\mathrm{market}(0) \, x, \; 
    x^{\top} P_\mathrm{market}(\nicefrac{1}{2}) \, x \} + c_1^2} 
\end{align*}
with
\begin{align*}
    P_\mathrm{market}(\jw\rho) = \left( \begin{array}{*5c}
1 & \jw\rho & \jw\rho & \jw\rho & \nicefrac{1}{4} \\
\jw\rho & 1 & \nicefrac{3}{4} & \nicefrac{3}{4} & \nicefrac{1}{4} \\
\jw\rho & \nicefrac{3}{4} & 1 & \nicefrac{1}{2} & \nicefrac{1}{4} \\
\jw\rho & \nicefrac{3}{4} & \nicefrac{1}{2} & 1 & \nicefrac{1}{4} \\
\nicefrac{1}{4} & \nicefrac{1}{4} & \nicefrac{1}{4} & \nicefrac{1}{4} & 1 \\
\end{array}\right)
\end{align*}
\kd{and $\rho \in \{0,\nicefrac{1}{2} \}.$
Being a correlation matrix, $P_\mathrm{market}(\rho)$ is positive semi-definite. 
%In particular, for $\rho=0$ and $\rho = \nicefrac{1}{2}$, the matrix $P_\mathrm{market}(\rho)$ has only positive eigenvalues, so $P_\mathrm{market}(\rho)$ is positive definite. 
Consequently, the square root in $f_\mathrm{market}$ is well defined. }
\hide{
where 
\begin{align*}
    f_z:\mathbb{R}^6 & \to \{0, \nicefrac{1}{2}\} \\
    x &\mapsto\left\{\begin{array}{ll}
    0 & x_\mathrm{yieldup} > x_\mathrm{yielddown} \\
    \frac{1}{2} & \, \textrm{else} \\
    \end{array}\right.
\end{align*}
}
}

%-------------------
% Constant risk
%-------------------
Market risk is then aggregated with 
\jw{other risks that are not affected by capital allocation, and the ratio of aggregated risk and own funds forms the solvency rate.}
\kd{All operations that are independent of the portfolio weights, i.e., independent of the optimization variables, are summarized by constants $c_2,\dots,c_4 > 0$ and $c_5 \in \R$. Note that the risk ratio to be built is hidden in the constants $c_2$ and $c_5$. Finally we obtain the following simple form of aggregation:}
\begin{align*}
f_\mathrm{\jw{constantrisks}}:\R\kd{_0^+}&\to\R \\
x & \mapsto c_2 \sqrt{x^2 + c_3 x + c_4} + c_5.
\end{align*}
\kd{The solvency ratio used as one of the objective functions is then obtained by composing the previously introduced functions:}
\kd{
\begin{align} \label{eq:solvency}
    f_\mathrm{solvencyratio}: [0,1]^n&\to\R \\ \notag
\omega& 
 \mapsto
f_\mathrm{constantrisks}(
f_\mathrm{market}(
f_\mathrm{aggregation}(
f_\mathrm{netrisk}(
\omega
)))).
\end{align}
}

%% file: application.tex
%---------------------------------
%
%---------------------------------
\section{Application to the strategic asset allocation}
\label{sec_impl}

\begin{figure}
    \centering
    \includegraphics[width = .8\textwidth]{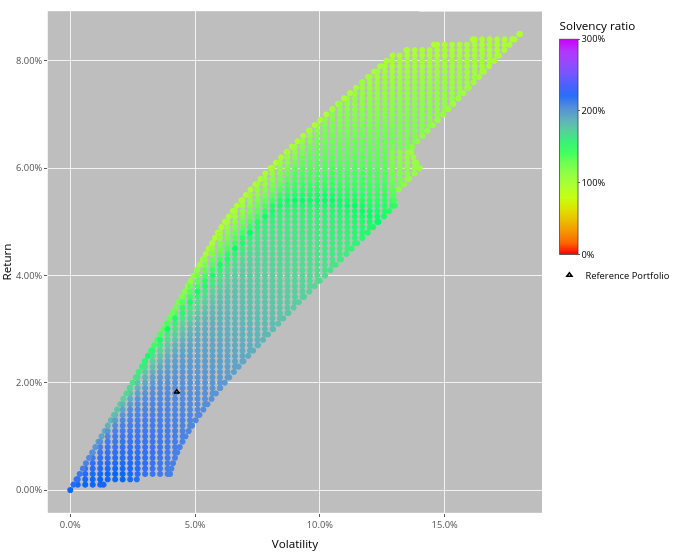} %scale=0.4
    \caption{Based on the 13 asset classes from Table~\ref{tab:assetclasses}, a discretization of feasible return-volatility combinations in the outcome space is computed. %In a second step, 
    The solvency ratio is maximized in all grid points.}
    \label{fig:solvency}
\end{figure}

\kd{In this section we solve the four-criteria optimization problem presented in Section~\ref{sec_model} with the help of the box algorithm described in Section~\ref{sec_mco}.
The presented test case is a 'near-real-world' case. As a basis we use the data of our industrial partner. However, for not revealing company-related secrets, we have to modify the data slightly. 
As a result, we obtain $13$ exemplary assets with their individual returns and volatilites as given in Table~\ref{tab:assetclasses}.  

Figure~\ref{fig:solvency} gives an impression of all feasible portfolios, depicted in the image space with respect to return and volatility. 
Note that no additional bounds on single assets or asset groups are active that would rule out highly unrealistic portfolios as, e.g., the portfolio in the lower left corner, which has a return and volatility of $0\%$, respectively, and refers to a $100\%$ investment into cash.
The shape of the feasible set in the image space typically resembles a flame. This also holds true for the real-world data.
Note that the upper left boundary of the feasible set represents the image of the portfolios that are Pareto optimal with respect to the two objectives return and volatility, hence, the portfolios that would have been obtained with Markowitz' bicriteria optimization. 

As a third dimension, we depict the \emph{solvency ratio}. 
Note that Figure~\ref{fig:solvency} does not show the result of a tricriteria optimization but that}
a single-objective optimization problem maximizing the solvency ratio is solved in every grid point, i.e., by restricting portfolio return and volatility to the respective values in the image space. 
We call the resulting portfolios 'solvency-optimal' in the following. 
The attained \jw{solvency ratios} are given \jw{by the} color of the grid points. 
Figure~\ref{fig:solvency} shows the general behavior we observed for the solvency ratio. The highest values are typically found in the lower left part, i.e., where return and volatility are rather small. 

The solvency-optimal
portfolios are typically extreme in the sense that they invest only in few asset classes.
\rg{An example for this behavior is shown in Table~\ref{tab:solvencyoptimalweights}, where the weights of the reference portfolio and a solvency-optimal portfolio with similar return and volatility are given.
}
\kd{(The reference portfolio is depicted as a black triangle in Figure~\ref{fig:solvency}, the solvency-optimal portfolio we consider lies next to it in south-west direction, so it has a slightly smaller value for return and volatility, respectively.)}

\begin{table}
\caption{Asset allocations of the reference portfolio and a solvency-optimal portfolio}
    \label{tab:solvencyoptimalweights}
    \begin{tabular}{l|r|r}
    \hline\noalign{\smallskip}
        \multicolumn{1}{c|}{Asset class $i$} & 
        \multicolumn{1}{c|}{Reference Portfolio} & 
        \multicolumn{1}{c}{Solv-opt Portfolio} \\
\noalign{\smallskip}\hline\noalign{\smallskip}
        Real Estate Germany & 5.98\% & 13.26\%\\
        Real Estate Intl.\ & 1.20\% & 0.00\%\\
        Equity Intl.\ Large Cap & 2.39\% & 14.86\%\\
        Equity Germany\ Large Cap & 15.55\% & 0.00\%\\
        Equity Intl.\ Small Cap & 0.60\% & 0.00\%\\
        Emerging Markets Equities & 0.60\% & 0.00\%\\
        Private Equity  & 0.12\% &0.00\%\\
        Government Debt & 29.90\% & 54.51\%\\
        Corporate Debt & 17.94\% & 0.00\%\\
        Infrastructure Finance & 0.60\% & 0.00\%\\
        Fixed Income & 4.78\% & 0.00\%\\
        Asset Backed Securities & 14.35\% & 0.00\%\\
        Cash & 5.98\% & 17.37\% \\ 
        \noalign{\smallskip}\hline\noalign{\smallskip}
        Return & 1.83\% & 1.80\% \\
        Volatility & 4.27\% & 4.20\% \\
        Solvency & 191.64\% & 206.28\% \\
                Distance & 0.00\% & 111.50\% \\
        \noalign{\smallskip}\hline
    \end{tabular}
\end{table}

It turns out that 
\rg{
the usability of such a solvency-optimal portfolio is} poor:
%meager: 
\kd{
Although it seems to be close to the reference portfolio, its allocation in the pre-image space differs considerably from the allocation of the reference portfolio. 
Indeed, evaluating \eqref{eq:distance} yields a value of $111.5 \%$ for the distance between the two portfolios. 
This shows the motivation for using an additional criterion that takes the distance to the reference portfolio into account without imposing hard constraints for the asset weights. 
This is discussed in the following. 
}

%---------------------------------
%
%---------------------------------
\subsection{\kd{Algorithmic details}}

In order to overcome %\rgout{this problem} 
the problem described above, we %\rgout{propose to }
consider an optimization problem with four objectives: 
\emph{return}, 
\emph{volatility}, 
\emph{solvency} and the 
\emph{distance to a reference portfolio}.

The multi-objective optimization algorithm is implemented in Java 8. The scalarizations are solved by invoking NLOpt which is a library available on  \url{http://github.com/stevengj/nlopt}. %\cite{nlopt}.
NLOpt offers a multitude of global and local optimization algorithms. We use their implementation of the Sequential Least Squares Programming (SLSQP) optimizer.

The graphical user interface is implemented in RShiny version 1.3.2. %(\cite{shiny}). 
\kd{There, } the user selects the objectives that should be included in the optimization.
Consequently, our framework also allows to  \kd{consider less objective functions.}

%---------------------------------
%
%---------------------------------
\subsection{Visualization of multi-objective portfolios}

\begin{figure}
    \centering
    \includegraphics[width = 0.9\textwidth]{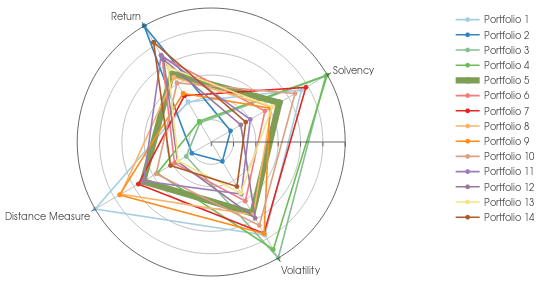} %scale=0.7
     \includegraphics[width = .9\textwidth]{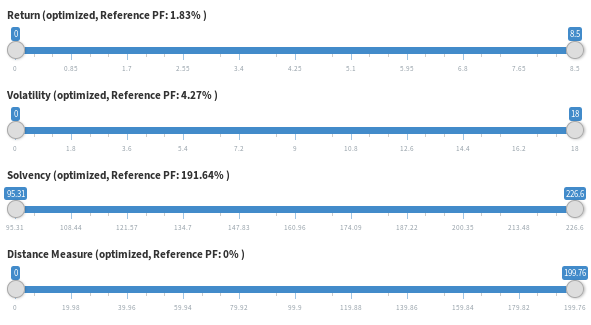} %scale=0.7
    \caption{Radar plot containing information regarding the four criteria return, volatility, solvency and distance to a reference portfolio. 
    Portfolio 5, which is the first compromise to be generated, is \jw{emphasized}. %due to Example \ref{ex:1}.
    The sliders below show the ranges of the four criteria, respectively. 
    }
    \label{fig:radarplot_1}
\end{figure}

All computed portfolios are presented to the user numerically and with the help of dedicated visualization techniques. In particular, we choose radar plots, which are one of the classical visualization approaches in multi-objective optimization, 
see, e.g., \cite{Miettinen2014} for a survey. 
An example is given in Figure \ref{fig:radarplot_1}.

Each portfolio is represented by a color. 
\kd{Since two of the considered objective functions are minimized and two are maximized, we unify the representation in the radar plot by inverting the minimized objective function values. Hence, for all objectives it holds that the more outer on the circle, the better the performance in the considered objective function (smaller in the minimization case and larger in the maximization case).}
For example, the dark blue portfolio in Figure~\ref{fig:radarplot_1} has the largest return of all generated portfolios while, e.g., the light blue portfolio \kd{has the smallest} distance measure. 

Together with the radar plot we offer sliders, see also Figure \ref{fig:radarplot_1}.
By moving the sliders, the nondominated points are filtered. 
Visually, the filtered nondominated points are grayed out in the radar plot.  
\kd{Since the reference portfolio plays an important role, its value in each of the considered objectives is additionally displayed in the title of the respective slider. Note that by definition, the reference portfolio has a distance measure of $0\%$ while all other portfolios might differ by a value between $0\%$ and $200\%$ from it. }

\begin{table}
\caption{Results of Algorithm~\ref{alg:box} when no further restrictions are active. The first four portfolios refer to the solutions defining the bounds of the initial box. The first portfolio is the reference portfolio.
}
\label{tab:radar}       % Give a unique label
\begin{tabular}{l|rrrr}
\hline\noalign{\smallskip}
PF & \multicolumn{1}{c}{return} & \multicolumn{1}{c}{volatility} & \multicolumn{1}{c}{solvency} & \multicolumn{1}{c}{distance} \\
     & \multicolumn{1}{c}{$[0,8.5\%]$} & \multicolumn{1}{c}{$[0, 18\%]$} & \multicolumn{1}{c}{$[95.31, 226.6\%]$} & \multicolumn{1}{c}{$[0, 199.76\%]$} \\
\noalign{\smallskip}\hline\noalign{\smallskip}
1 & $1.83\%$& $4.27\%$& $191.64\%$& $0.00\%$\\
 2 & $8.50\%$ & $18.00\%$& $95.3\% $& $199.76\%$ \\
 3 & $0.00\%$ & $0.00\%$ & $224.37\%$& $188.04\%$\\ 
 4  & $0.12\%$ & $1.58\%$ & $226.60\%$ & $128.23\%$ \\ \noalign{\smallskip}\hline\noalign{\smallskip}
 5  & $4.29\%$ & $8.31\%$ & $161.57\%$ & $98.95\%$ \\
 6  & $5.79\%$ & $10.60\%$ & $142.04\%$ & $163.87\%$ \\
7 &	    $2.37\%$ &	$4.59\%$ &	$197.54\%$ & $89.27\%$ \\
8 &	    $3.99\%$ &	$8.04\%$ &	$151.43\%$ &	$52.97\%$ \\
9 &     $2.57\%$ &  $4.54\% $ & $148.70\% $ & $50.35\%$ \\
10  &   $3.49\%$ & $6.08\%$ & $182.91\% $ & $125.88\%$ \\
11  &   $5.93\%$ & $11.90\%$ & $122.15\% $ & $100.05\%$ \\
12  &   $5.54\%$ & $7.44\%$ & $109.24\% $ & $170.04\%$ \\
13  &   $4.88\%$ & $12.09\%$ & $151.85\% $ & $171.60\%$ \\
14  &   $7.07\%$ & $13.32\%$ & $116.01\% $ & $155.60\%$ \\
\noalign{\smallskip}\hline
\end{tabular}
\end{table}

\kd{

\subsection{Computational results with four objectives}

In the following we present two use cases. The first shows the application of Algorithm~\ref{alg:box} to Problem~\eqref{eq:prob}
without further restrictions.

\begin{example}\label{ex:1}
In Figure \ref{fig:radarplot_1} \kd{and Table~\ref{tab:radar}} we present
the results of Algorithm~\ref{alg:box} when applied to Problem~\eqref{eq:prob} for the input shown in Table~\ref{tab:assetclasses}.
The four additional portfolios that are computed in the beginning to determine the bounds of the starting box are also depicted and denoted as Portfolios~1--4. 
Here, the initial bounds are $l_0=(0\%,0\%,95.31\%,0\%)$ and $u_0=(8.5\%,18\%,226.6\%, 199.76\%)$. 
Portfolio~5 is the first that is computed in the initial search box. 
We emphasize this portfolio in Figure~\ref{fig:radarplot_1} to highlight that it roughly lies in the middle of the bounds of all criteria.
\kd{This shows the advantage of using a weighted Tchebycheff scalarization which typically generates solutions lying in the middle of the considered box.}
The algorithm now proceeds in decomposing the initial search box with respect to this point into 
new hyperboxes. It selects one of the boxes \kd{according to \eqref{eq:boxselect}, i.e., it refines the box with the largest minimal edge,} and searches for a solution in it. 
In our example, this 
results in Portfolio~6. 
\kd{The algorithm ends with the generation of Portfolio~14. 

When considering the outcomes depicted in Figure~\ref{fig:radarplot_1} and Table~\ref{tab:radar}, 
we notice that already the ten generated intermediate portfolios $5-14$ cover the initial search region 
$[0\%,8.5\%] \times [0\%, 18\%] \times [95.31\%, 226.6\%] \times [0\%, 199.76\%]$ entirely, 
in the sense that the intermediate portfolios have different well-distributed values over all components. 
This is one of the main advantages of Algorithm~\ref{alg:box} over existing approaches like the one used in \cite{Xidonas2018}: 
We can specify any desired number of iterations (and, thus, portfolios to be generated) and for any such input the algorithm will produce a representation covering the entire initial search region 'uniformly'. 
In contrast, grid-based approaches only allow to specify the number of intervals $q_i$, into which the range of each objective $i=2,\dots,m$ is equally divided. 
The $q_i$ intervals result in $q_i - 1$ intermediate equidistant grid points, in total $(q_2+1) \cdot (q_3+1) \dots (q_m+1)$ scalarizations are solved. 
For having a representation of approximately $10$ portfolios for the given four-criteria problem, the user would have to choose $q_i\in \{1,2\}$ for $i=2,3,4$, 
resulting in $8$, $12$, $18$ or $27$ iterations. 
Note that in the first case, no intermediate grid point would be generated at all, in the second case only one intermediate grid point would have been generated in only one objective.  
This demonstrates that for an increasing number of objectives, many more iterations are required to achieve a similar representation than with our approach.
}
\end{example}

\begin{figure}[h]
    \centering
    %\subfloat[The distance measure is additionally limited to a maximum of $50\%$. \label{subfig:radar3}]
    \includegraphics[width=.8\textwidth]{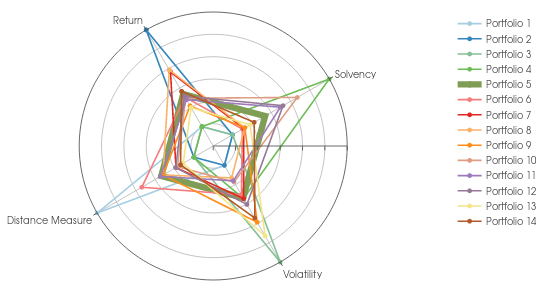} \hfill 
    \includegraphics[width=.8\textwidth]{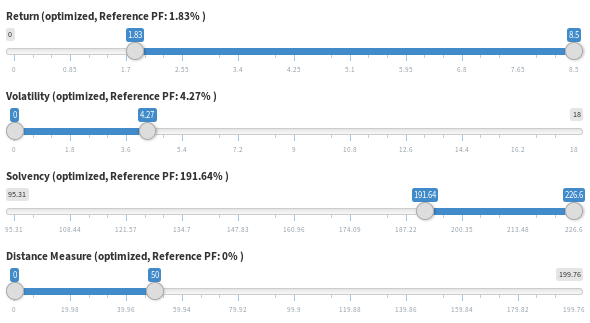}
    \caption{Visualization of Example~\ref{ex:3}: The sliders corresponding to return, volatility and solvency are restricted to values that are all at least as good as the reference portfolio. The distance measure is additionally limited to $50\%$.
    }
    \label{fig:radarplot_2}
\end{figure}

\kd{So far, we have not restricted Problem~\eqref{eq:prob} further. However, as discussed in the beginning of Section~\ref{sec_impl}, the user typically has a strong interest in low transaction costs. 
There are two ways to achieve this goal. One is to use hard bounds on assets or asset groups. This approach requires a lot of additional input and probably also a lot of fine-tuning until a satisfying setting is found. Here, we propose another way which only needs one figure to be specified, namely the maximum distance from the reference portfolio. 

\begin{example} \label{ex:3}
We restrict now the distance measure to $50\%$. 
Furthermore, we bound the other criteria to enforce that the generated outcomes are better than the reference portfolio. In particular, we impose the additional constraint that  
return has to be larger than $1.83\%$, 
volatility smaller than $4.27\%$ 
and solvency larger than $191.64\%$.
The generated portfolios are depicted in Figure~\ref{fig:radarplot_2}, their objective values are given in Table~\ref{tab:radar50dist}. 
Note that the new restrictions have further effects on the bounds of the other criteria, as can be seen from the first four generated portfolios. 
Hence, the initial search region is now given by 
$[1.83\%, 2.33\%] \times [3.37\%, 4.27\%] \times [191.64\%, 201.57\%] \times [0\%, 50\%]$.
Again, we set the number of iterations to $10$. 
The algorithm generates $10$ intermediate portfolios that are, as shown in Table~\ref{tab:radar50dist}, distributed over the initial search region.

We now come back to the issue discussed at the beginning of Section~\ref{sec_impl}, where we selected a portfolio close to the reference portfolio with respect to return and volatility and maximized solvency. As shown in Table~\ref{tab:solvencyoptimalweights}, the solvency-optimal portfolio turned out to have an unexpectedly high distance measure of $111.5\%$. 
Since we now take the distance to the reference portfolio as one of the objectives into account, we expect that the allocations of the resulting portfolios differ much less from the reference portfolio. As an example, we have a closer look at 
Portfolio~4 from Table~\ref{tab:radar50dist}, which has the same return and a similar volatility and solvency as the reference portfolio.
In Table~\ref{tab:weights50dist} we show the corresponding allocation variables and compare them to the ones of the reference portfolio. A graphical version of the same information is provided in Figure~\ref{fig:detail_scr_3}. 
While Portfolio~4 still invests into less assets than the reference portfolio, its diversity is considerably better than the portfolio generated without the distance criterion. In this way, the user can direct the search quickly to interesting portfolios without the burden to find good bounds for the individual assets. 
\end{example}
}

\begin{table}
\caption{Results of Algorithm~\ref{alg:box} when distance is restricted to $50\%$ and all other criteria are restricted to the values of the reference portfolio. The first four portfolios refer to the solutions defining the bounds of the initial box.
The first portfolio is the reference portfolio. 
}
\label{tab:radar50dist}       % Give a unique label
\begin{tabular}{l|rrrr}
\hline\noalign{\smallskip}
PF & \multicolumn{1}{c}{return} & \multicolumn{1}{c}{volatility} & \multicolumn{1}{c}{solvency} & \multicolumn{1}{c}{distance} \\
     & \multicolumn{1}{c}{$[1.83,2.33\%]$} & \multicolumn{1}{c}{$[3.37, 4.27\%]$} & \multicolumn{1}{c}{$[191.64, 201.57\%]$} & \multicolumn{1}{c}{$[0, 50\%]$} \\
\noalign{\smallskip}\hline\noalign{\smallskip}
1   & $1.83\% $ & $	4.27\% $ & $	191.64\% $ & $	0.00\% $ \\
2   & $2.33\% $ & $	4.27\% $ & $	191.64\% $ & $	50.00\% $ \\
3   & $1.83\% $ & $	3.37\% $ & $	191.64\% $ & $	50.00\% $ \\
4   & $1.83\% $ & $	3.96\% $ & $	201.57\% $ & $	50.00\% $ \\
\noalign{\smallskip}\hline\noalign{\smallskip}
5   & $2.00\% $ & $	3.97\% $ & $	194.98\% $ & $	33.25\% $ \\
6   & $1.98\% $ & $	4.00\% $ & $	192.66\% $ & $	23.18\% $ \\
7   & $	2.11\% $ & $	3.96\% $ & $	192.79\% $ & $	40.77\% $ \\
8   &	$2.12\% $ & $	4.15\% $ & $	192.90\% $ & $	34.13\% $ \\
9   & $1.94\% $ & $	3.74\% $ & $	192.91\% $ & $	34.10\% $\\
10  & $1.98\% $ & $	4.11\% $ & $	198.29\% $ & $	41.59\% $\\
11  &	$1.97\% $ & $	4.13\% $ & $	196.55\% $ & $	32.83\% $\\
12  &	$1.99\% $ & $	3.91\% $ & $	196.82\% $ & $	40.66\% $\\
13  &	$1.93\% $ & $	3.61\% $ & $	193.44\% $ & $	44.49\% $\\
14  &	$2.01\% $ & $	3.78\% $ & $	193.87\% $ & $	43.17\%$\\
\noalign{\smallskip}\hline
\end{tabular}
\end{table}

\begin{figure}
    \subfloat[On the left, the reference portfolio and, on the right, Portfolio~4 of Table~\ref{tab:radar50dist}. 
    ]
    {
    \includegraphics[width=.98\textwidth]{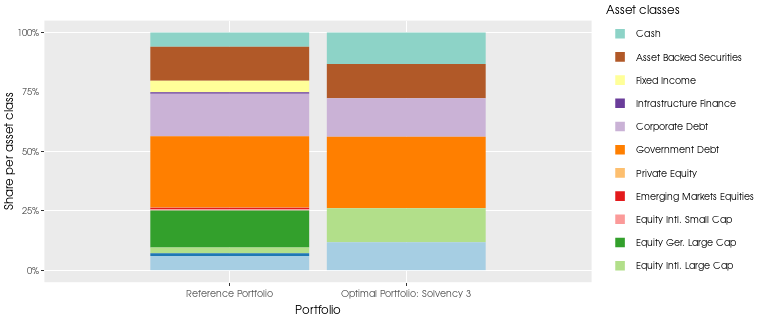} 
    }

    \subfloat[For each asset class the difference between the two portfolios is made visible. \label{subfig:scr3horiz}]
    {
    \includegraphics[width=.98\textwidth]{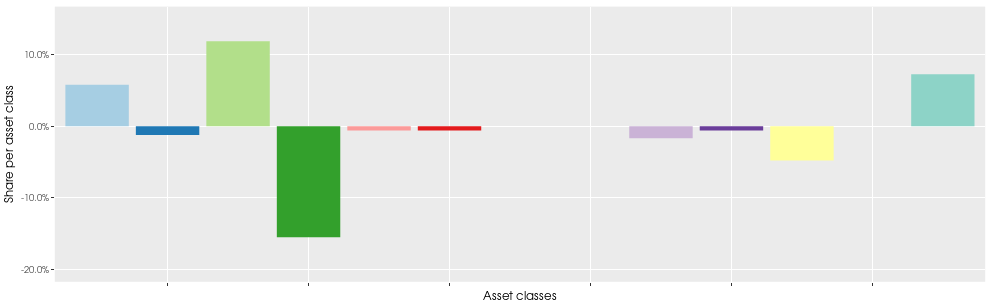}
    }

    \caption{Graphical comparison of the composition of the reference portfolio and Portfolio~4 of Table~\ref{tab:radar50dist}
    }
    \label{fig:detail_scr_3}
\end{figure}

\begin{table}
    \caption{Asset allocations of the reference portfolio and a portfolio with distance restricted to $50\%$}
    \label{tab:weights50dist}
    \begin{tabular}{l|r|r}
    \hline\noalign{\smallskip}
        \multicolumn{1}{c|}{Asset class $i$} & 
        \multicolumn{1}{c|}{Reference Portfolio} & 
        \multicolumn{1}{c}{Portfolio~4 of Table~\ref{tab:radar50dist}} \\
\noalign{\smallskip}\hline\noalign{\smallskip}
        Real Estate Germany & 5.98\% & 11.79\%\\
        Real Estate Intl.\ & 1.20\% & 0.00\%\\
        Equity Intl.\ Large Cap & 2.39\% & 14.30\%\\
        Equity Germany\ Large Cap & 15.55\% & 0.00\%\\
        Equity Intl.\ Small Cap & 0.60\% & 0.00\%\\
        Emerging Markets Equities & 0.60\% & 0.00\%\\
        Private Equity  & 0.12\% &0.12\%\\
        Government Debt & 29.90\% & 29.90\%\\
        Corporate Debt & 17.94\% & 16.27\%\\
        Infrastructure Finance & 0.60\% & 0.00\%\\
        Fixed Income & 4.78\% & 0.00\%\\
        Asset Backed Securities & 14.35\% & 14.35\%\\
        Cash & 5.98\% & 13.27\% \\ 
        \noalign{\smallskip}\hline\noalign{\smallskip}
        Return & 1.83\% & 1.83\% \\
        Volatility & 4.27\% & 3.96\% \\
        Solvency & 191.64\% & 201.57\% \\
        Distance & 0.00\% & 50.00\% \\
        \noalign{\smallskip}\hline
    \end{tabular}
\end{table}

}